\documentclass{article}

\usepackage{subfig}
\usepackage{natbib}

\usepackage{arxiv}

\usepackage[utf8]{inputenc} 
\usepackage[T1]{fontenc}    
\usepackage{hyperref}       
\usepackage{url}            
\usepackage{booktabs}       
\usepackage{amsfonts}       
\usepackage{nicefrac}       
\usepackage{microtype}      
\usepackage{lipsum}
\usepackage{graphicx}
\graphicspath{ {./images/} }

\title{Imperceptible Gaze Guidance Through Ocularity in Virtual Reality}

\author{
 Virmarie Maquiling\\
  Human-Centered Technologies for Learning\\
  Technical University of Munich\\
  Munich Germany \\ 
  \texttt{virmarie.maquiling@tum.de} \\
   \And
 Li Zhaoping \\
  Max Planck Institute for Biological Cybernetics\\
  University of Tübingen\\
  Tübingen Germany\\
  \texttt{zhaoping.li@tuebingen.mpg.de} \\
  \And
 Enkelejda Kasneci \\
  Human-Centered Technologies for Learning\\
  Technical University of Munich\\
  Munich Germany \\ 
  \texttt{enkelejda.kasneci@tum.de} \\
}

\begin{document}
\maketitle
\begin{abstract}
We introduce to VR a novel imperceptible gaze guidance technique from a recent discovery that human gaze can be attracted to a cue that contrasts from the background in its perceptually non-distinctive ocularity, defined as the relative difference between inputs to the two eyes. This cue pops out in the saliency map in the primary visual cortex without being overtly visible. We tested this method in an odd-one-out visual search task using eye tracking with 15 participants in VR. When the target was rendered as an ocularity singleton, participants' gaze was drawn to the target faster. Conversely, when a background object served as the ocularity singleton, it distracted gaze from the target. Since ocularity is nearly imperceptible, our method maintains user immersion while guiding attention without noticeable scene alterations and can render object's depth in 3D scenes, creating new possibilities for immersive user experience across diverse VR applications. 
\end{abstract}

\begin{figure}
    \centering
  \includegraphics[width=0.8\textwidth]{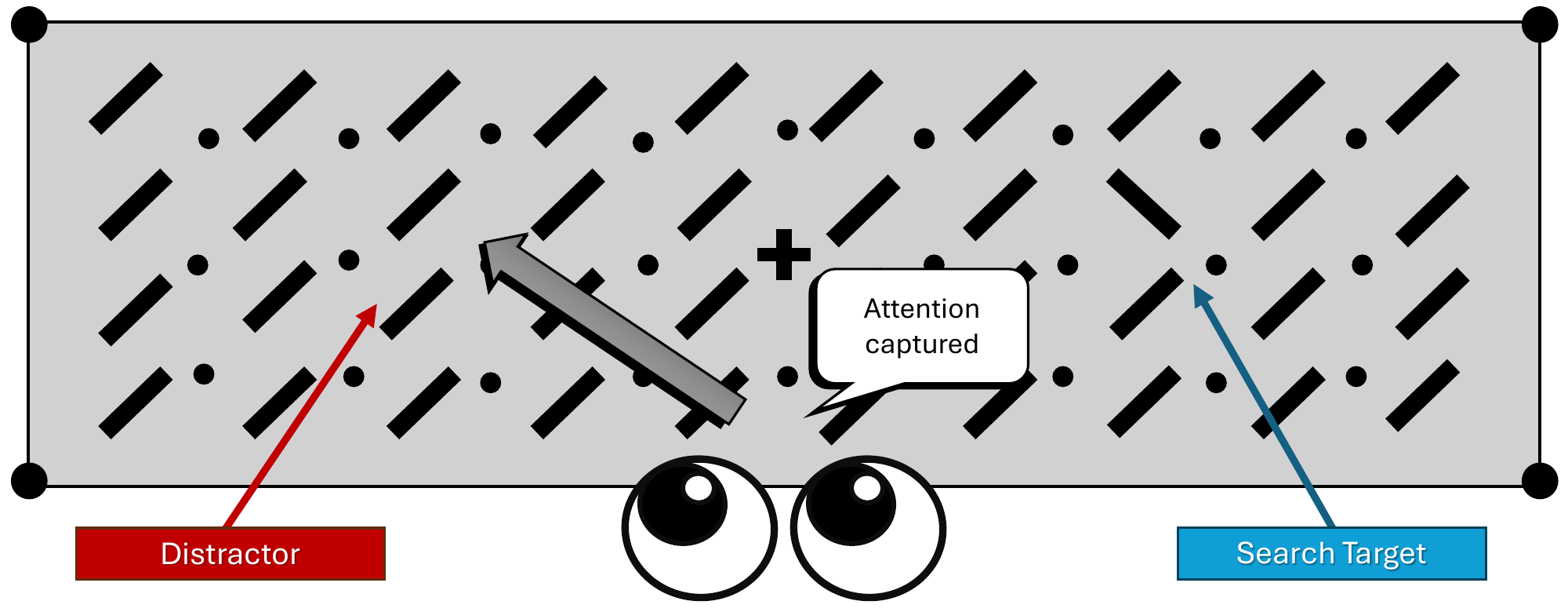} 
  \caption{We conducted an odd-one-out visual search task where the participant is asked to find the orientation singleton (a unique bar facing a different direction among uniformly oriented bars). In this case, a random distractor is rendered as an ocularity feature singleton which causes the participant to instinctively look at the distractor instead of finding the target. Ocularity features deal with individual ocular inputs to each eye (e.g. present in one eye while invisible to another). Such an alteration is not perceivable to the beholder as they only see the combined image from both eyes}
  \label{fig:teaser}
\end{figure}

\section{Introduction}
Extended reality (XR) technologies, such as Meta Quest, PlayStation VR, and Apple Vision Pro, have surged in popularity, bringing with them a critical need to maintain high levels of user immersion. A key challenge across applications—from training to medical simulations and gaming—is guiding user attention without disrupting this immersive experience. Explicit visual cues, such as arrows \cite{lin2017tell, gruenefeld2018flyingarrow}, swarms \cite{lange2020hivefive, kergassner2024hivefive360}, and spatial blurs \cite{hata2016visual}, are often used to direct attention effectively. However, these overt cues modify the user’s visual environment, add visual clutter \cite{sutton2022look}, can potentially feel overpatronizing to the user, and can break user immersion \cite{grogorick2020subtle}. Addressing these issues, subtle alternatives have been explored, such as those proposed by \citeauthor{bailey2009subtle}~[\citeyear{bailey2009subtle}], \citeauthor{grogorick2017subtle}~[\citeyear{grogorick2017subtle}], and \citeauthor{erickson2022analysis}~[\citeyear{erickson2022analysis}], which rely on peripheral visual modulations or differences in information presented to each eye. While these approaches are less intrusive, they still alter the scene and risk being noticed by the user. This raises an important question: can we guide attention in VR without any perceptible change to the visual environment? Partially addressing this, \citeauthor{krekhov2018deadeye}[\citeyear{krekhov2018deadeye}] introduced a technique that renders objects visible to only one eye, but this compromises depth perception, limiting thus its effectiveness. 

To go further, we turn to the concept of saliency, which has been extensively studied in vision science. In this context, we define saliency as the measure of how much a region of an image stands out from its surroundings, drawing attention in a bottom-up, exogenous manner \cite{zhaoping2022parallel}. The V1 Saliency Hypothesis (V1SH) \cite{li2002saliency, zhaoping2014v1} posits that the brain’s primary visual cortex (V1) generates a saliency map, guiding reflexive gaze shifts based on visual information. This is significant because certain features can capture attention without conscious perception, making them ideal candidates for imperceptible gaze guidance. Recent research has shown that ocularity singletons—ocularity being defined as contrast differences between inputs to the left and right eyes—can trigger exogenous gaze shifts via iso-feature suppression (i.e. the suppression of neuronal response to a feature, such as color or orientation, when similar features are present nearby) \cite{knierim1992neuronal, jones2001surround, wachtler2003representation}.  All while remaining perceptually invisible to the user \cite{zhaoping2008attention, zhaoping2012gaze, zhaoping2018ocularity}. This suggests a promising avenue for imperceptible gaze guidance, but its application in immersive XR environments remains unexplored.

In this paper, we investigate the potential of ocularity modulation to imperceptibly guide attention in a VR environment. We hypothesize that this method can direct user attention without any perceivable change to the visual scene, offering a novel form of gaze guidance that preserves immersion. Our contributions are twofold: First, we introduce a new approach to gaze guidance using ocularity as a means to attract gaze. Second, we provide empirical validation of the V1 Saliency Hypothesis in a VR setting, presenting eye tracking results from an odd-one-out search task that demonstrate its efficacy in real-world applications beyond controlled laboratory environments.

\section{Related Works}

\subsection{Gaze Guidance Techniques}
Over the years, various gaze guidance techniques have been developed, broadly categorized into two types. The first are overt techniques which use visual cues that are deliberately noticeable, typically relying on global image modulation or in-game elements designed to prompt a saccadic movement. Examples of these methods include the use of dots \cite{jarodzka2013learning}, arrows \cite{gruenefeld2018flyingarrow}, swarms \cite{lange2020hivefive, kergassner2024hivefive360}, semantic depth-of-field \cite{kosara2002focus+} and many more. 

Going the opposite direction, subtle techniques aim to guide attention in a less intrusive manner, using cues designed to be minimally perceptible or visually unobtrusive. This is highly important in many VR applications where immersion is paramount. One of the best known is \citeauthor{bailey2009subtle}'s Subtle Gaze Guidance [\citeyear{bailey2009subtle}], which exploits the low acuity of the peripheral vision and the fact that humans are incapable of processing visual information during a saccadic movement (saccadic masking). Their method involved applying subtle image-space modulation, particularly using luminance as well as warm-cool modulation performed over pixels in the target region, in the user's periphery and using eye tracking technology to monitor saccadic velocity, ensuring that the modulation terminates before the modulated area enters the user's foveal vision. This was later extended to VR by \citeauthor{grogorick2017subtle}[\citeyear{grogorick2017subtle}], adding stimulus shape variation and dynamic stimulus positioning. Other works include saliency modulation \cite{sutton2022look, sato2016sensing}, subtle spatial blur \cite{hata2016visual}, and temporal luminance modulation \cite{grogorick2018comparison}. 

More recent works have also explored binocular rivalry--a phenomenon where the brain alternates between two different images presented to each eye, rather than merging them into a single coherent perception \cite{blake2001primer}--as a way to capture attention. \citeauthor{grogorick2020stereo}~[\citeyear{grogorick2020stereo}] exploited binocular rivalry through stereo inverse brightness modulation. The method involves presenting conflicting brightness information to each eye, inducing a perceptual conflict that attracts attention without overtly altering the scene. \citeauthor{erickson2022analysis}~[\citeyear{erickson2022analysis}] investigated the effectiveness of color-based dichoptic visual cues in optical see-through augmented reality (OST AR) displays. DeadEye \cite{krekhov2018deadeye, krekhov2019deadeye} leverages dichoptic presentation, showing different stimuli to each eye to guide attention without altering the visual properties of the target object. Specifically, they attract attention by rendering the object only to one eye, arguing that this induces binocular rivalry, where the visual system automatically detects the conflict between the two images. This perceptual mismatch causes the target object to``pop out'' preattentively, without consciously altering the visual scene, thus preserving the overall integrity of the visualization while effectively guiding the viewer's gaze.

\subsection{V1 Saliency Hypothesis}
The V1 Saliency Hypothesis \cite{li2002saliency, zhaoping2014v1} proposes that the primary visual cortex (V1) generates a bottom-up saliency map of the visual field from retinal inputs to guide visual attention exogenously, independent of higher-level cognitive processes This saliency map is created from neurons in the V1 region that respond more strongly to unique or rare features in the visual environment, creating peaks on the saliency map, highlighting areas in the visual field that are likely to attract attention. One intriguing finding from this research is the attention-grabbing effect of ocularity singletons--stimuli with unique ocularity, such as a bar visible to the right eye among similar bars visible to the left. These ocularity singletons can capture attention even when not consciously perceived \cite{zhaoping2008attention, zhaoping2012gaze}, supporting the idea that V1 can drive attention based on bottom-up processing alone, without higher-level cognitive involvement. Ocularity is defined as the relative difference between input contrasts to the left $C_L$ and right $C_R$ eye, where $C = (L-L_0) / max(L-L_0)$ is the normalized luminance $L$ of an input item where $L_0$ is the background luminance \cite{zhaoping2018ocularity}. In short, a binocular (ocularly balanced) item has an ocularity of $O=0$ while a monocular item has an ocularity of $O=\pm1$. Further, a right-eye dominant monocular item has a negative ocularity while a left-eye dominant item has positive ocularity.   

Given that ocularity singletons depend on disparities in visual inputs between the eyes, they present a novel opportunity for gaze guidance in virtual reality (VR) systems, which use stereoscopic displays to simulate depth. While existing stereo-based gaze guidance is often framed through the lens of binocular rivalry, we approach it from a different perspective, focusing on ocularity itself as a feature that can attract gaze. Although both concepts share parallels, binocular rivalry involves competition between monocular neurons, resulting in alternating dominance between the two eyes' perceptions \cite{blake2001primer}. In contrast, we view ocularity through the framework of V1SH, treating it as a controllable feature that can directly affect an object's saliency without modifying the fused image perceived by the observer while preserving depth perception \cite{zhaoping2018ocularity}.

\section{Methodology}
\subsection{Experimental Setup and Procedure}
\begin{figure}
    \centering
    \includegraphics[width=\linewidth]{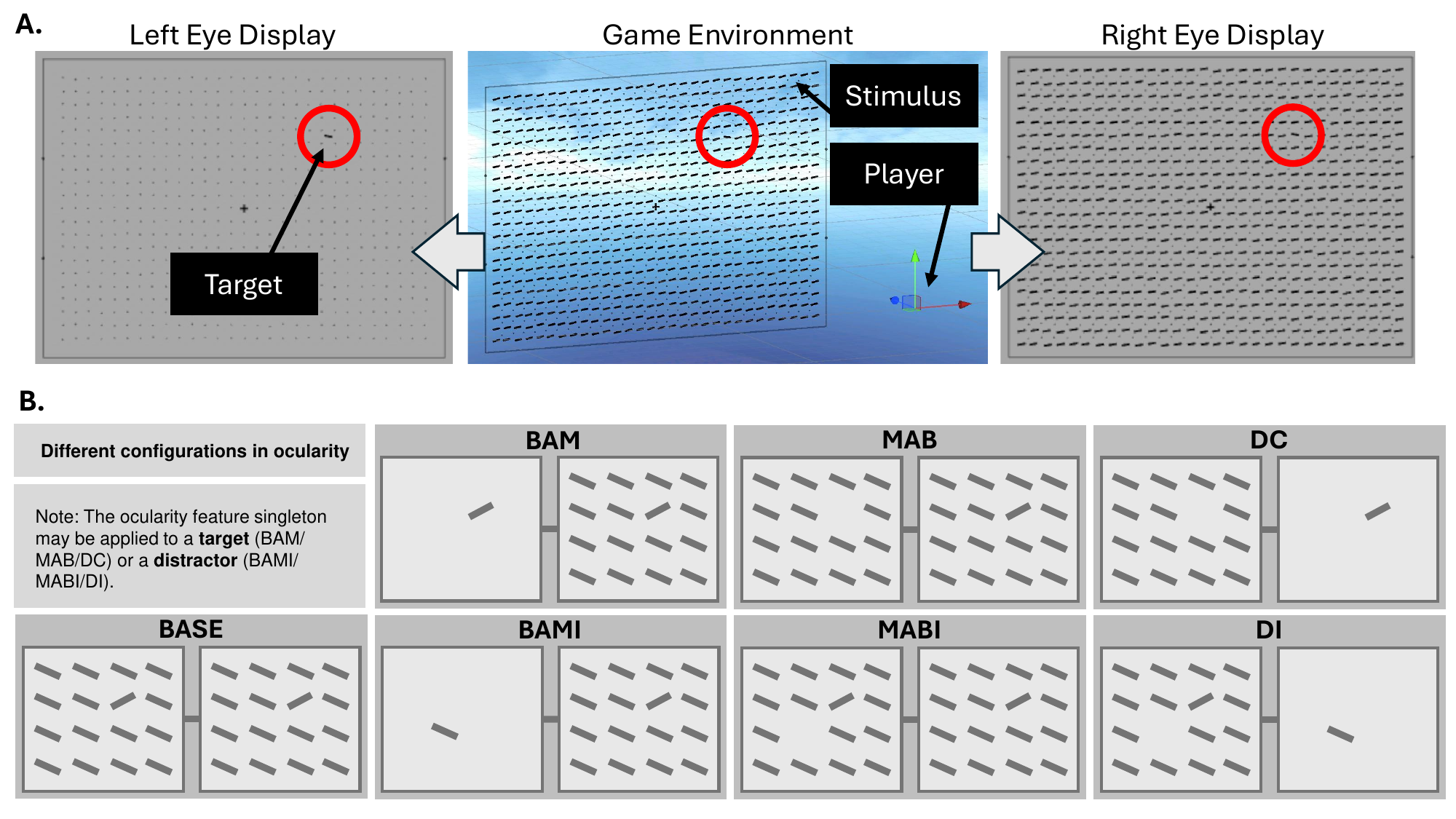}
    \caption{A. Middle: The game environment on Unity showing the Player facing the stimulus. Specifically, the stimulus  mimicks a 2D grid composed of 3D objects floating in space. Head rotation is disabled to simplify the eye tracking data collection process. The gaze vector is provided by the eye tracking module and is translated to 2D screen coordinates (XY) in pixels. The player's head movement in-game was disabled. Left and Right: The game as seen from the left and right displays, respectively. In this example, a trial with a BAM condition is shown where non-targets are displayed to the right eye while the target (encircled in red in the three images) is shown to both eyes. B. Different configurations in ocularity used in the task. Each image pair represent a simplified example of what the participant sees in VR through the left and right eye display. In this example, all the non-targets are displayed to the right eye. In the experiment, this is randomly selected. Similarly, the orientation of the bars are randomly selected. Depending on the condition, the ocularity singleton can either be a target or a distractor.}
    \label{fig:environment}
\end{figure}

To analyze the saliency of ocular feature singletons, we conducted an odd-one-out visual search task, akin to the experiments conducted by \citeauthor{zhaoping2008attention}~[\citeyear{zhaoping2008attention, zhaoping2018ocularity}]. In this task, participants were instructed to locate a visually unique object within a grid of similar-looking objects, specifically, they need to find the orientation singleton--a single bar facing a certain direction among similar-looking bars facing the other direction) as exemplified in Fig. \ref{fig:environment}. The grid comprised of 30 columns and 22 rows of uniformly oriented bars, each rotated 10 degrees upward from the horizontal axis. The bars were randomly raised from either the left or right side. In order to anchor the vergence, the grid was enclosed by a rectangular outline, while each grid cell was marked by a circular dot at its bottom-right corner. To avoid a wallpaper effect \cite{mckee2007wallpaper, nakamizo1999subjective}, the anchoring dots are slightly shifted vertically or horizontally by a very small amount. Written and graphical instructions were presented at the start of the experiment, instructing participants to press the left control key if the unique target appeared on the left side of the grid (relative to the fixation cross) or the right control key if it was on the right side. Between each trial, the fixation cross at the center of the grid flickered for 1.17 s before the bars appeared, during which, the participant is instructed to look at the cross. Upon key press, the target flickered for 375 ms functioning as visual feedback to confirm the response, thereby reducing potential frustration. This is then followed by a short delay to give the participant an opportunity to freely view the grid. No specific instructions were given to the participant during this delay. 

Within the grid, specific locations equidistant from the fixation cross were selected as potential target locations. The target position for each trial was randomly assigned among these candidates. To simplify our experiment, we chose the highest possible contrast for ocularity, along with complete ocular balance, setting $O = \{-1, 0, 1\}$. In doing so, the bars were only rendered in one of two ways: a) binocular, in which the bar was visible for both eyes, and b) monocular, in which the bar was visible to only one eye. Whether the monocular object is visible to the right or the left eye is randomly selected. 

Each trial was assigned one of seven randomly chosen conditions: 
\begin{itemize}
    \item BASE: all bars are binocular
    \item BAM (Binocular Among Monoculars): binocular target among monocular non-targets 
    \item BAMI (Binocular Among Monoculars - Incongruent): binocular non-target (distractor) among monocular target and non-targets 
    \item MAB (Monocular Among Binoculars): monocular target among binocular non-targets 
    \item MABI (Monocular Among Binoculars - Incongruent): monocular distractor among binocular targets and non-targets 
    \item DC (Dichoptic Congruent): monocular target present in one eye while monocular nontargets present in the other eye  
    \item DI (Dichoptic Incongruent): monocular distractor present in one eye while monocular target and nontargets present in the other eye
\end{itemize}

The distractor was a non-target bar programmed to appear somewhere on the opposite side of the grid from the target. This design allowed us to systematically explore how different visual conditions influence the saliency and detection speed of ocular feature singletons in a controlled VR environment. Furthermore, incorporating distractors as ocularity singletons, enabled us to observe how well it can interfere top-down tasks (searching for the orientation singleton).

\subsection{Participants}
We collected data from 17 participants (10 females and 7 males), aged 22-36 years (M=28.1, SD=5.4), most of whom were university students with normal or corrected-to-normal vision, as confirmed through a pre-study visual screening. None of the participants had prior knowledge of ocularity singletons. After providing informed consent, participants were briefed on the study’s purpose, procedures, and potential risks. The study received full ethical approval from the relevant ethics committee. This approval covered all aspects of participant recruitment, data collection, and the protection of participants' rights and privacy throughout the VR-based gaze guidance tasks. Participants were also informed of their right to withdraw at any time without consequence. Prior to the experiment, each participant completed a questionnaire assessing their prior experience with VR, any known visual impairments, and susceptibility to motion sickness. Following the experiment, participants were asked to complete a brief post-questionnaire, providing feedback on their experience and offering any additional comments or observations.
 
We conducted the experiment using a Varjo VR-3 (Model HS-6) headset with eye tracking capabilities. Eye tracking is recorded at 200 Hz. Trials that were recorded with less than 70\% sampling ratio were excluded from the analysis. A 5-point calibration process is conducted before the start of the game. Additionally, a 4-point calibration is conducted in-game (by asking the participant to gaze at a pink dot that appears at the 4 corners of the grid) as an additional quality check. After visual validation of the quality of the eye tracker recordings, two participants were excluded from the study due to a consistently high presence of noise (possibly due to incorrectly detected pupils). The results presented in this paper were derived from the remaining 15 participants.

\subsection{Evaluation}

We calculated trial duration (reaction time) as a primary measure of saliency \cite{wolfe1996visual}, defined as the time it took for the participant to complete the task (pressing the button), with shorter reaction times indicating higher target saliency and easier detection. In addition, we measured several key aspects of participants’ eye movement behaviors. These measures include: 
\begin{itemize}
    \item \textbf{Average Fixation count}: The number of fixations made during the search task. This is typically counted per area of interest (AOI) \cite{goldberg1999computer}. In our case, we consider the target side (the side of the grid containing the target), the background side (the side not containing the target) and the overall grid as AOIs. In the results presented below, this metric counts the overall fixations across the entire grid.  A high number of fixations indicates inefficient search \cite{goldberg1999computer, sharafi2015systematic}, suggesting that the target was less salient or that the participant had difficulty locating it. In contrast, lower fixation counts are typically associated with more efficient search behavior and higher target saliency.
    \item \textbf{First Fixation Probability}: Here, we shift our focus on the target and background AOIs. First fixations often is associated with high visual saliency \cite{lohse1997consumer, bialkova2011efficient, van2015you}. This measure represents the likelihood that the participant's first fixation will land on the target side of the grid rather than the background side as the gaze moves away from the center region. The ``center region'' is defined as an area extending 10 pixels from the fixation cross along the x-axis.
    \item \textbf{Fixation probability on the target and background side}: The probability that participants fixated on the target side of the grid versus the background side. A higher fixation probability on the target side suggests that participants spent more time fixating on the target, while a higher fixation probability on the background side indicates that participants spent more time searching in the wrong areas.
    \item \textbf{Scanpath Width}: This refers to the overall breadth of the scanpath, indicating how much the gaze spreads across the visual field. A wider scanpath suggests more dispersed fixations, while a narrower scanpath indicates more concentrated fixations within a smaller area of the grid.
\end{itemize}

Additionally, we measured the player's accuracy for each condition, defined as the proportion of correct responses (i.e., when participants correctly identified the target location) relative to the total number of trials in that condition.

\section{Results}

\begin{figure}[htbp]
    \centering
    \subfloat[Mean Trial Duration]{\includegraphics[width=0.5\textwidth]{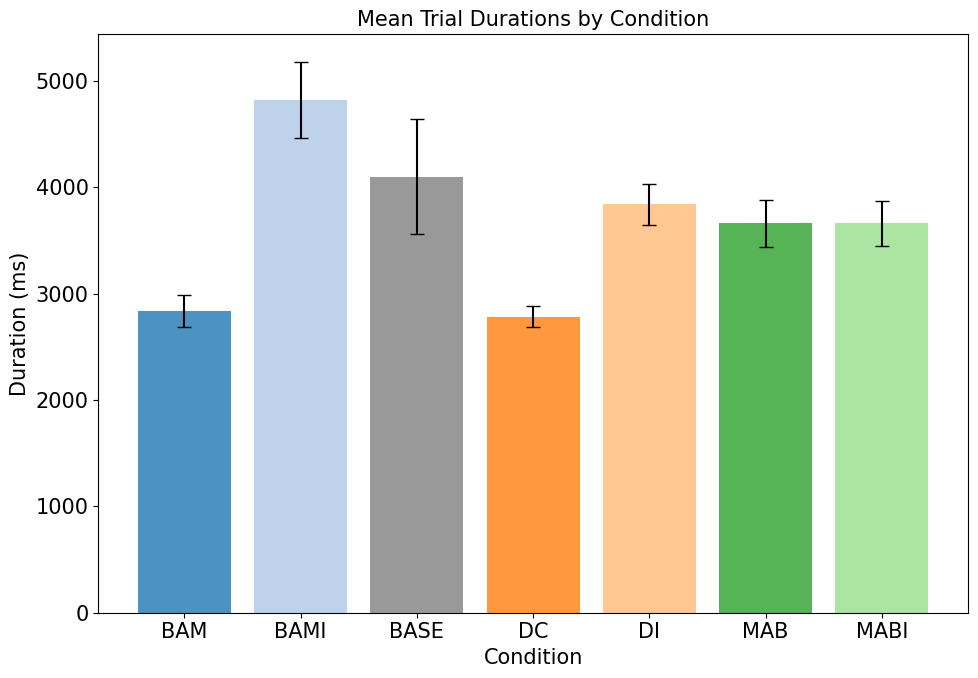}} \hfill
    \subfloat[Player Accuracy]{\includegraphics[width=0.5\textwidth]{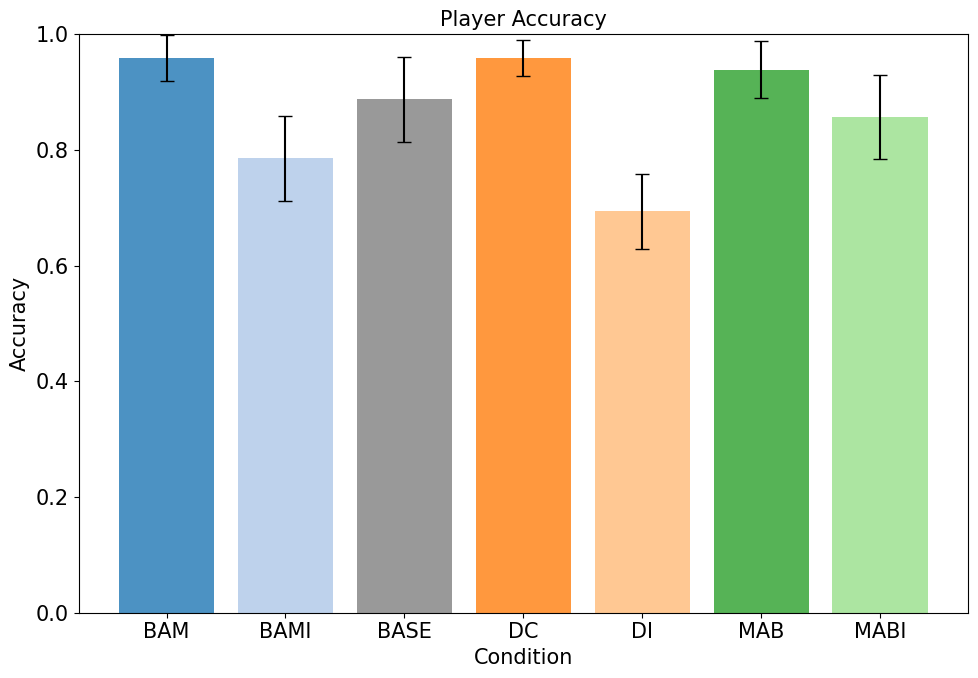}} \hfill
    \subfloat[Scanpath Width]{\includegraphics[width=0.5\textwidth]{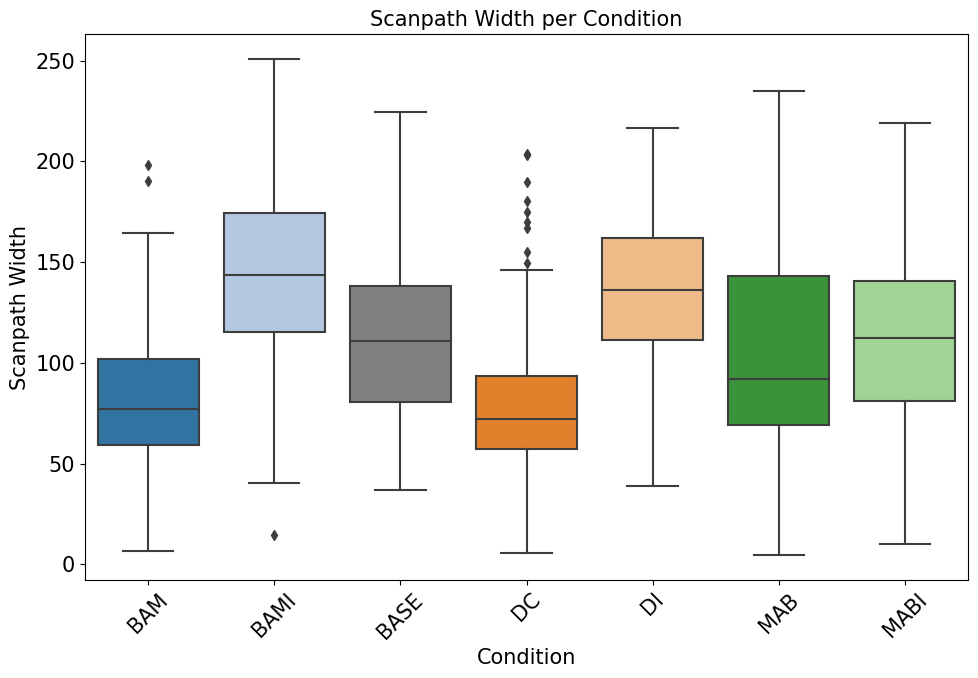}} \hfill
    \subfloat[Average Fixation Count]{\includegraphics[width=0.5\textwidth]{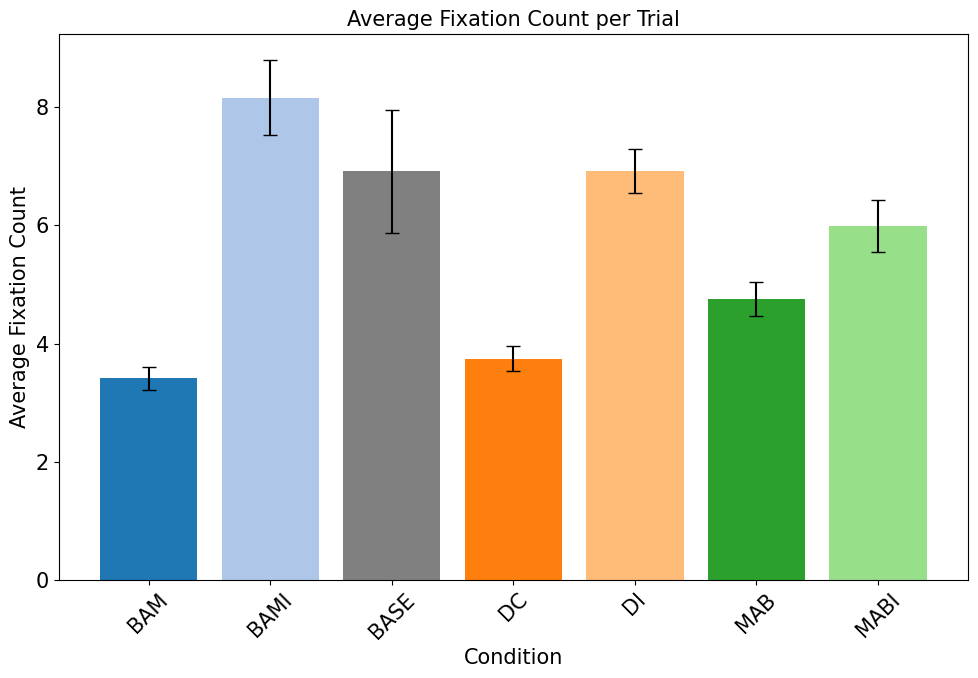}} \hfill
    \subfloat[First Fixation Probability]{\includegraphics[width=0.5\textwidth]{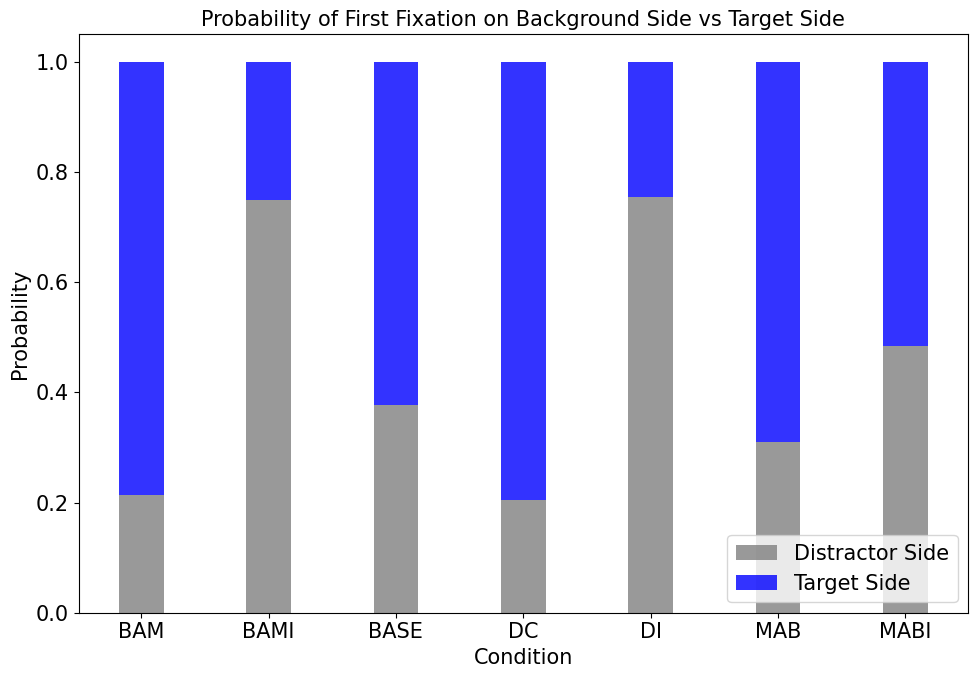}} \hfill    
    \subfloat[Average Fixation Probability: Target vs Background]{\includegraphics[width=0.5\textwidth]{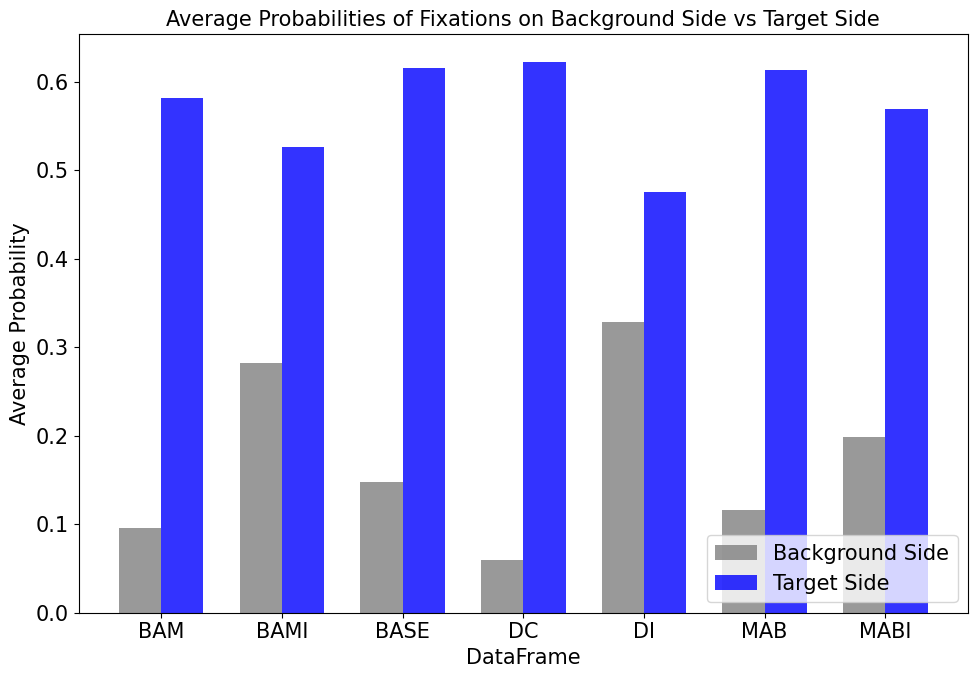}} 
    \caption{Mean trial duration, accuracy and eye tracking results across different conditions.}
    \label{fig:ETresults}
\end{figure}

All statistical comparisons across conditions were conducted using a Kruskal-Wallis test to assess overall differences, followed by post hoc Dunn’s tests with Bonferroni correction for pairwise comparisons. The Kruskal-Wallis test showed significant differences between conditions (p < 0.001). Dunn's test revealed that the pair BAM and DC as well as BAMI and DI conditions consistently showed no statistical differences between each other. Meanwhile, both MAB and MABI conditions showed no statistical difference with BASE. The results are visualized in Fig. \ref{fig:ETresults}. 

\subsection{Trial Duration and Player Accuracy}
Trial durations were shortest in the BAM and DC conditions (both 2.7 s), indicating that the target’s high saliency in both conditions enabled participants to quickly identify the target and complete the task. Dunn’s test confirmed that both BAM and DC durations were significantly shorter than all other conditions (all p < 0.001). In contrast, the BAMI condition had the longest average duration (4.8 s), suggesting that the distractor’s saliency caused confusion and prolonged the search process. This pattern highlights how a salient distractor can substantially disrupt search efficiency by drawing attention away from the target.
The highest average accuracy scores were observed in the BAM (96.2\%), DC (96.2\%), and MAB (94.3\%) conditions, which were target-salient. In contrast, distractor-salient conditions—BAMI (78.1\%, p = 0.32 vs. BAM), DI (70.5\%, p < 0.001 for all comparisons), and MABI (86.7\%, p = 0.032 vs. BAM)—showed significantly lower accuracy, suggesting that distractor salience impaired performance by diverting attention away from the target.

\subsection{Eye Tracking Metrics} 
Scanpath width was significantly narrower in the BAM (mean: 89.2 pixels) and DC (mean: 89.6 pixels) conditions, suggesting that participants’ attention was more focused when the target was an ocularity singleton. In contrast, scanpath width was much broader in the BAMI (mean: 151.2 pixels) and DI (mean: 142.9 pixels) conditions, indicating more dispersed search behavior and greater confusion when the distractor was the ocularity singleton. Both BAM and DC had significantly narrower scanpaths compared to BAMI, DI, and BASE (all p < 0.001). 
Participants made fewer fixations in BAM (mean: 3.41) and DC (mean: 3.74) conditions, reflecting a more efficient visual search. In contrast, BAMI and DI conditions had significantly higher fixation counts (mean: 8.16 and 6.92, respectively), indicating that participants struggled to locate the target. Dunn’s post hoc test showed that fixation counts in both BAM and DC were, again, significantly lower than BAMI, DI, and BASE (all p < 0.001).  
Analysis of first fixation probabilities showed that participants were significantly more likely to fixate on the target side first in the BAM (81.3\%) and DC (80.0\%) conditions (p < 0.001 compared to BAMI, DI, and MABI), indicating that when the target was an ocularity singleton, participants' attention was drawn to it immediately. In contrast, first fixation probabilities were significantly higher on the distractor side in the BAMI (79.6\%) and DI (80.0\%) conditions (p < 0.001 compared to BAM, BASE, DC, MAB, and MABI), suggesting that a salient distractor initially captured attention, leading to a delayed or less direct search process. 
Overall, the average fixation probabilities showed a significantly higher likelihood of fixating on the target side in BAM and DC (81.3\% and 80.0\% respectively, p < 0.001 compared to BAMI, DI, and MABI and p < 0.01 compared to BASE), with lower probabilities in the distractor-salient conditions like BAMI and DI (79.6\% and 80.0\% respectively, p < 0.001 compared to BAM, DI, BASE, MAB, and MABI), where participants spent more time fixating on the background or distractor side. This pattern suggests that target-salient conditions facilitated more focused and efficient search behavior, while distractor-salient conditions disrupted this process, increasing the likelihood of initial fixations on the distractor side.

\subsection{Qualitative Analysis} 
\begin{figure}[htbp]
    \centering
    \subfloat{\includegraphics[width=0.45\textwidth]{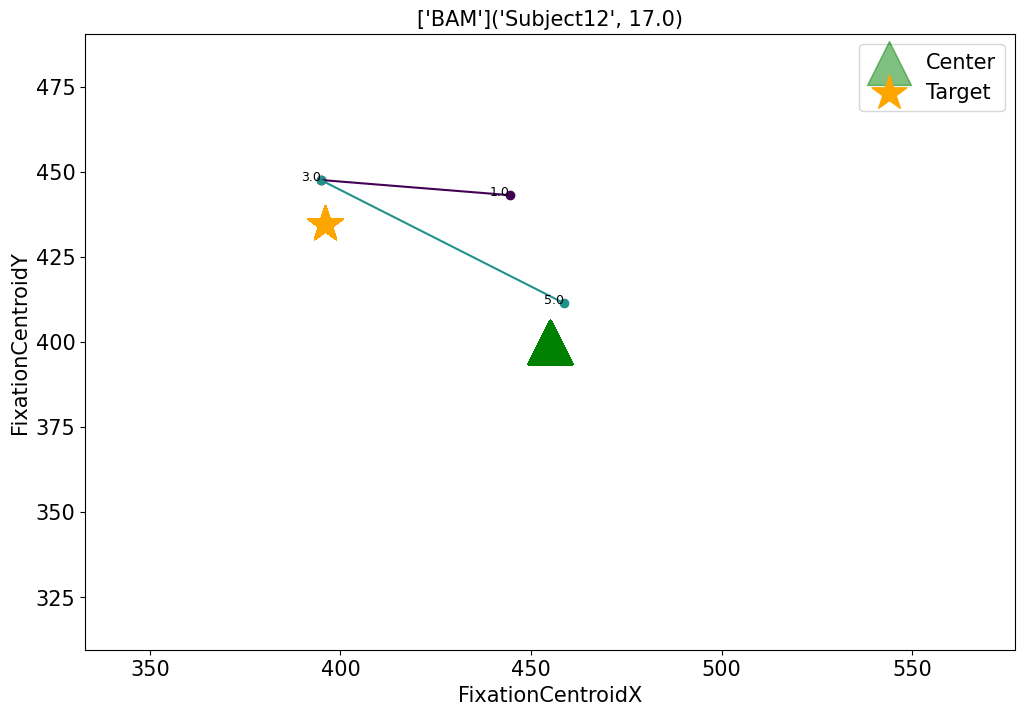}} \hfill
    \subfloat{\includegraphics[width=0.45\textwidth]{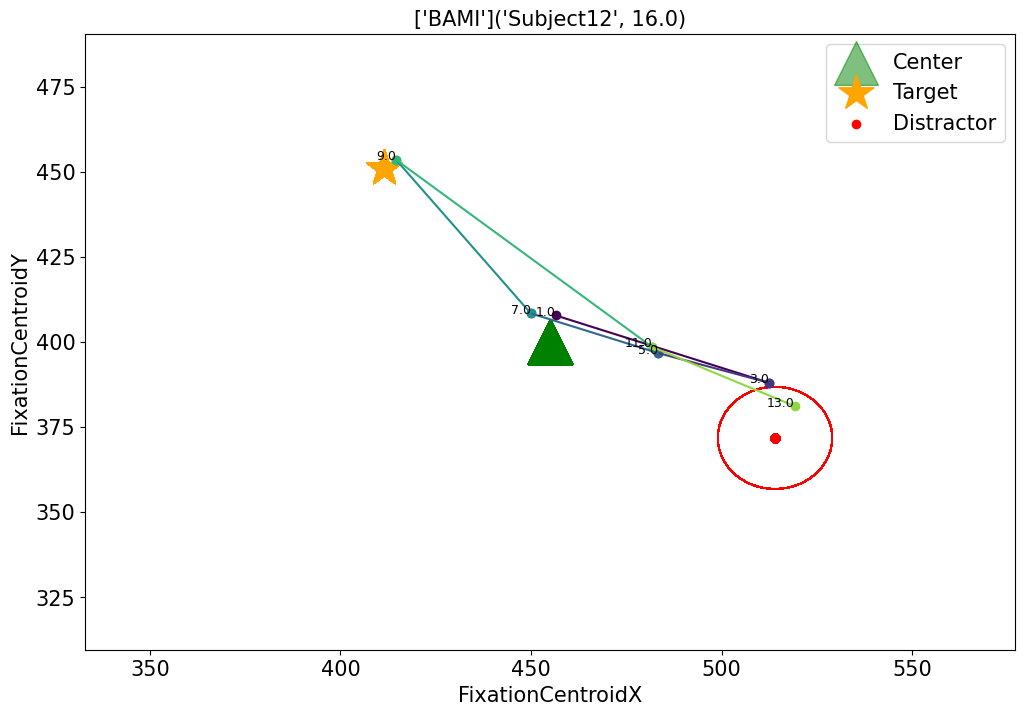}} \hfill
    \subfloat{\includegraphics[width=0.45\textwidth]{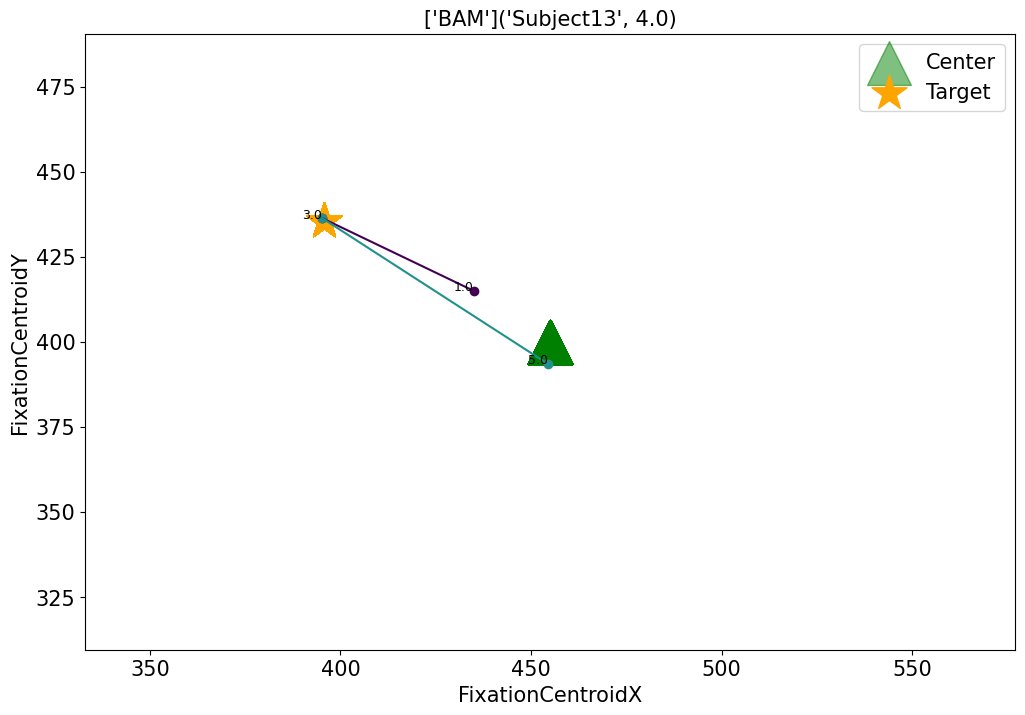}} \hfill
    \subfloat{\includegraphics[width=0.45\textwidth]{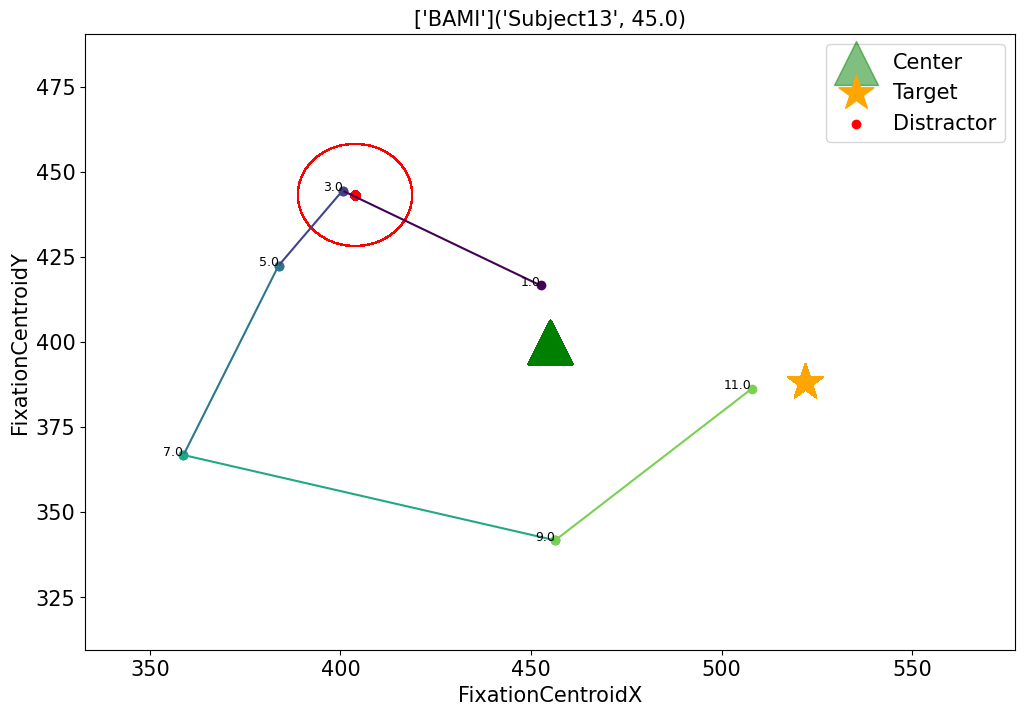}} \hfill
    \subfloat{\includegraphics[width=0.45\textwidth]{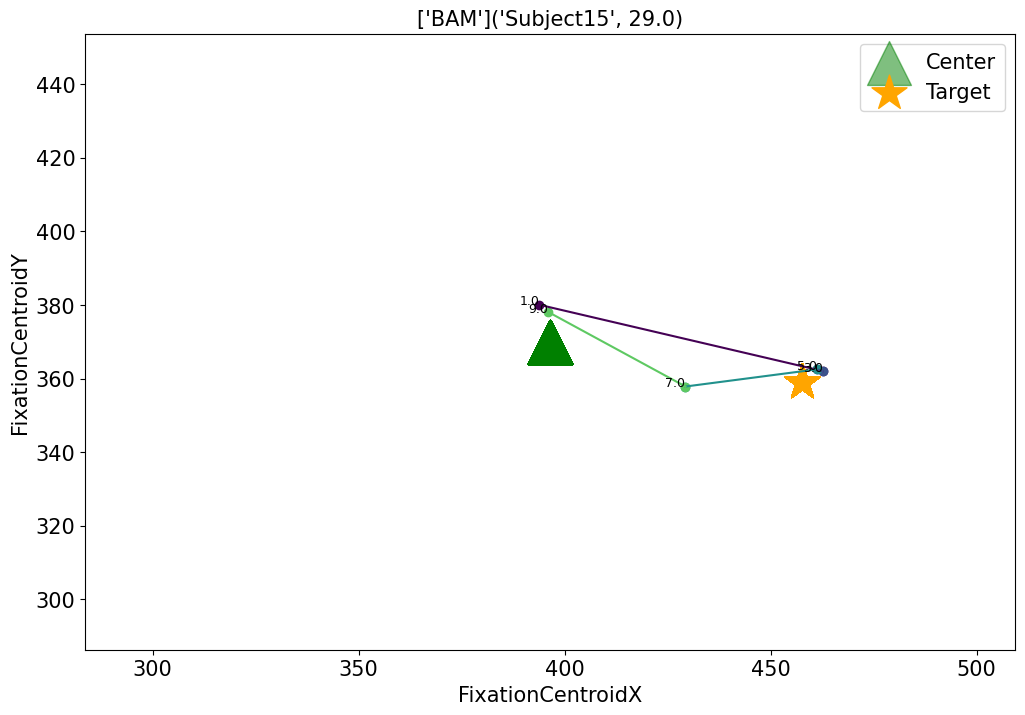}} \hfill
    \subfloat{\includegraphics[width=0.45\textwidth]{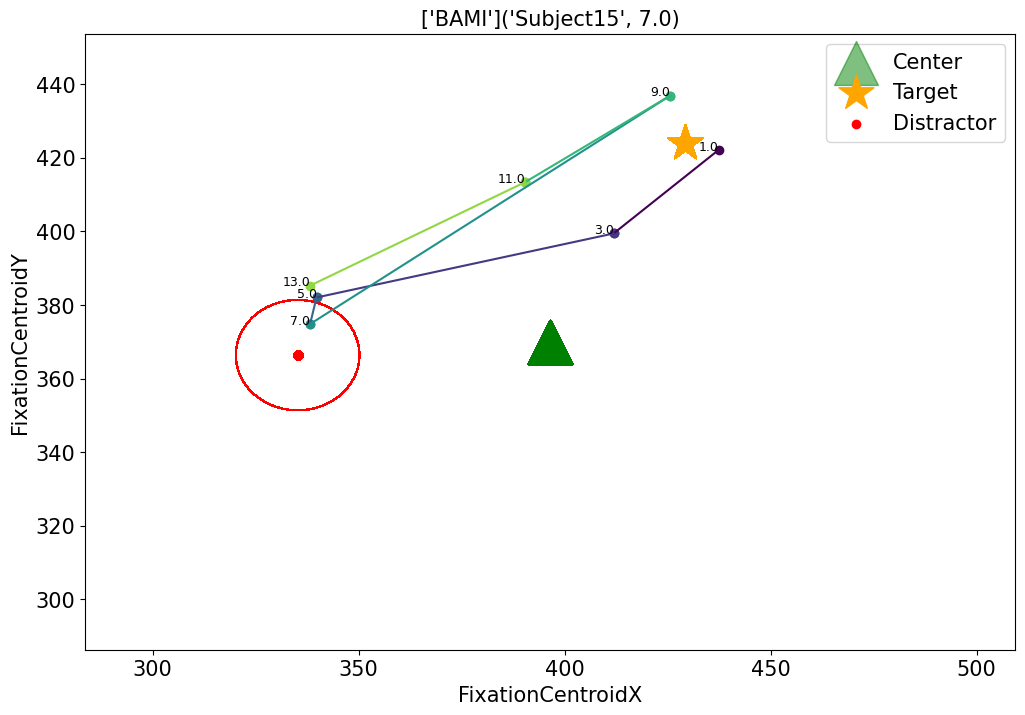}} \hfill
    \subfloat{\includegraphics[width=0.45\textwidth]{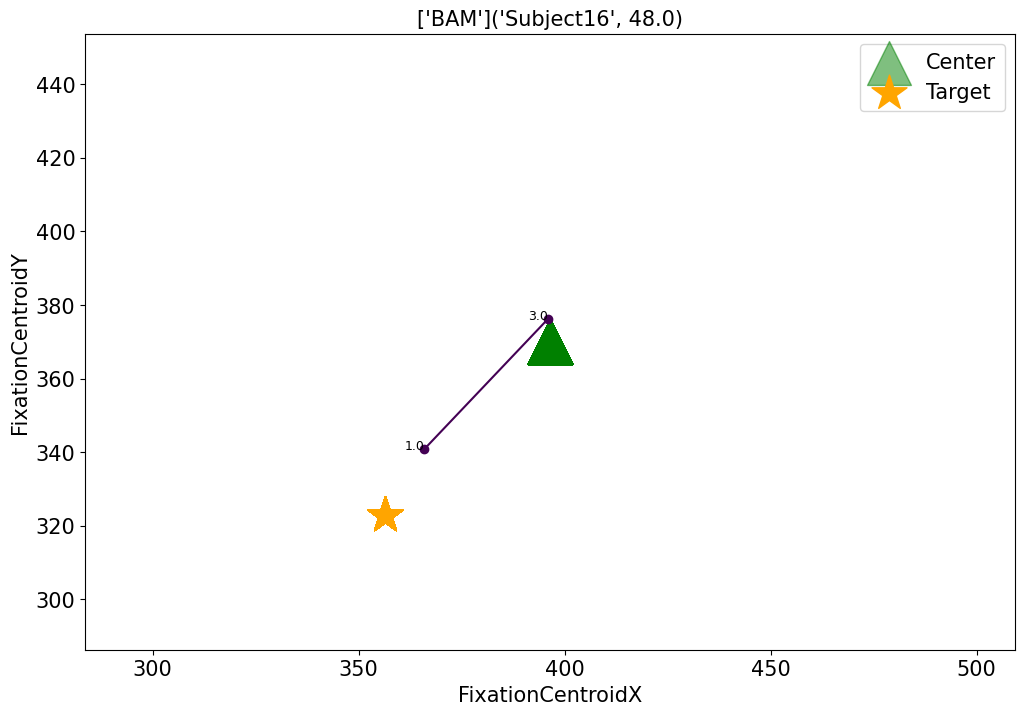}}\hfill
    \subfloat{\includegraphics[width=0.45\textwidth]{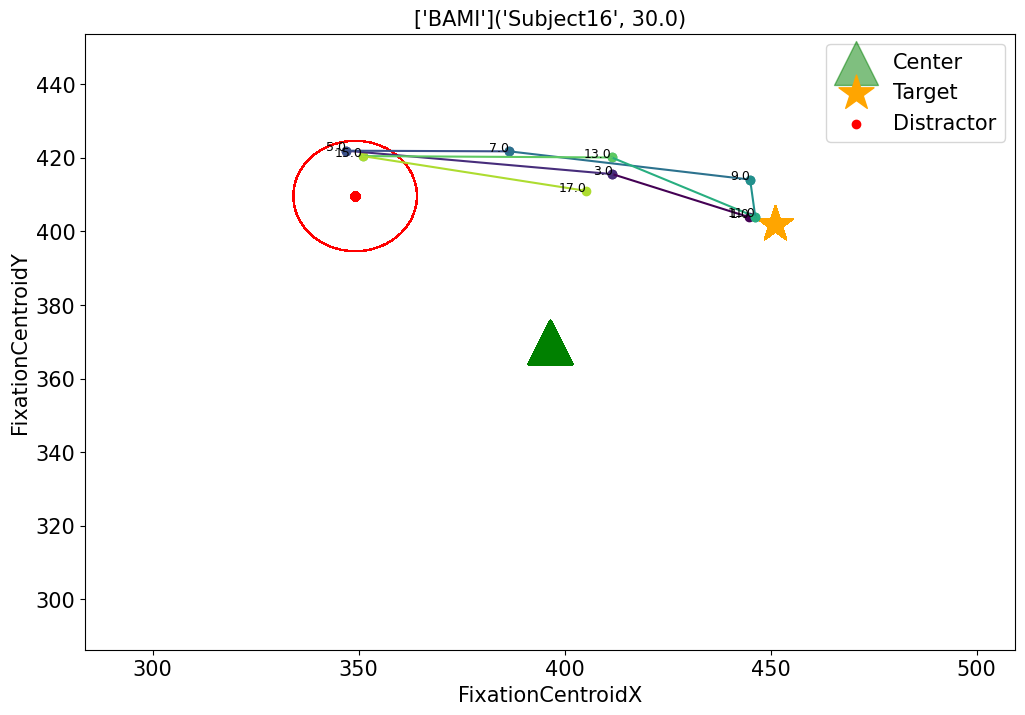}}
    \caption{Contrastive search behavior as exemplified by sample scanpaths taken from BAM (left) and BAMI (right): the search ended quickly for salient target conditions with the first fixation landing usually on the target (star) itself while the participant produced more dispersed fixations in the presence of salient distractors (circle) where the distractor itself is the first to be gazed at by the subject. The scanpath color indicates temporal order, with darker shades representing earlier fixations.}
    \label{fig:scanpaths}
\end{figure}

Upon analyzing the scanpaths, we quickly observed two distinct scanning behaviors based on whether the ocularity singleton was the target or the distractor. When the target was salient (i.e., in conditions BAM, DC, MAB), the scanpaths were notably shorter, with viewers often making immediate saccades toward the target, bypassing other items entirely. In contrast, when the distractor was salient (i.e., in conditions BAMI, DI, MABI), the scanpaths were wider and more dispersed. Viewers frequently shifted their gaze back and forth between the target and distractor, as though uncertain about the target’s location. Representative scanpaths illustrating these behaviors are shown in Fig. \ref{fig:scanpaths}. This pattern aligns with our findings above, highlighting the confusion introduced by salient distractors and their impact on the search task.

After the experiment, the participants were given the option to freely write down any feedback regarding their experience. Some participants described unique perceptions of the target/distractor, noting that it ``stood out'' in certain trials and was even detectable without direct eye movement. Additionally, while some described it as ``relaxing", ``easy", or ``fun'', a few participants mentioned difficulties with binocular fusion, particularly those unaccustomed to contact lenses or with significant differences between their eyes. Interestingly, some participants also reported a learning effect, with targets becoming easier to locate as they grew accustomed to the visual setup, suggesting an adaptation phase that may impact search efficiency. These findings not only highlight the effects of the ocularity singleton but also suggest potential refinements, such as incorporating practice trials or an additional segment for targeted feedback on specific conditions.

\section{Discussion}
\begin{figure}
    \centering
    \includegraphics[width=\linewidth]{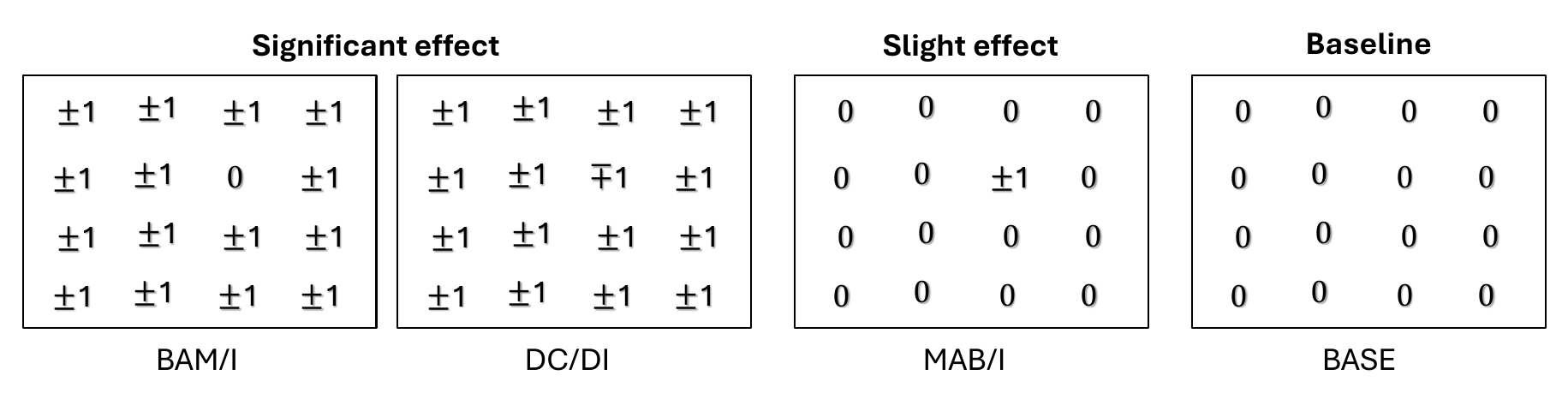}
    \caption{A comparison of BAM/I, DC/I, MAB/I, and BASE conditions in terms of ocular contrast where each item in the grid is represented by its corresponding ocularity $O$. Each condition shows a simplified $4\times4$ grid where the ocularity singleton (in all conditions apart from BASE) is found on the 3rd column in row 2. BAM/I and DC/DI conditions showed significant effect on the participant's search behavior while MAB/I demonstrated a small change in search behavior.}
    \label{fig:ocularity}
\end{figure}

The results demonstrate that different configurations of ocularity significantly affect search behavior. Participants frequently fixated on the distractor, even when it lacked perceptual features that would typically attract attention. This indicates that ocularity singletons can override top-down tasks, as participants were actively searching for the orientation singleton but were still drawn to the distractor despite it being visually indistinctive. Some participants even reported mistakenly identifying the distractor as the target, underscoring the powerful draw of ocularity features in capturing attention. %
Conditions like the BAM/BAMI and DC/DI appeared to be the most salient. The BAM and DC conditions facilitated instantaneous target detection, as participants quickly located the target. In contrast, BAMI and DC induced confusion, with the distractor becoming highly salient and diverting attention away from the target. This finding supports the idea that modifying ocularity can significantly influence attention, particularly when a distractor becomes more salient than the target. For a visual comparison between conditions in terms of ocularity, refer to Fig. \ref{fig:ocularity}.

While MAB did not show statistical significance over the baseline, this may align with previous findings by \citeauthor{zhaoping2018ocularity}~[\citeyear{zhaoping2018ocularity}], which suggested that in scenes mostly containing ocularly balanced items, a unique visual item that has a high ocular imbalance (e.g. MAB) induces avoidance behavior. Although this was not directly studied in our experiment, several participants made comments similar to \citeauthor{zhaoping2018ocularity}'s earlier observations regarding their experience in the MAB condition. \citeauthor{zhaoping2018ocularity}~[\citeyear{zhaoping2018ocularity}] proposed that this avoidance behavior can be overridden by timing the ocularity feature to last just long enough to trigger a saccadic response but short enough to prevent the avoidance behavior—this could be an area for future exploration, potentially through the use of a gaze-contingent window. Interestingly, the MABI condition did not significantly differ from the baseline, likely because the distractor's presumed ``lustre'' led participants to avoid looking in the area where it was present.

\subsection{Limitations, Open Questions, and Future Work}

Our study has certain limitations. Conducted in a controlled VR setting with a simple visual search task, the experiments disabled head rotation and only included participants with normal or corrected-to-normal vision. Thus, we cannot generalize these findings to individuals with visual impairments. Additionally, our study did not directly compare this method with existing gaze guidance techniques, as our primary goal was to provide VR-based evidence of the gaze-guiding effects of ocularity singletons. %
Moreover, the ocularity in our experiment was limited to three discrete values ($O=\{-1, 0, 1\}$). An area for further exploration is the impact of weakly balanced ocularity singletons, where $-1 < O < 1$ and $O \ne 0$ (e.g., a singleton rendered at full opacity to one eye and half opacity to the other). This approach would allow us to directly control the saliency of 3D objects while keeping them visible to both eyes. 

Several questions emerged from this initial study. Some participants noted difficulty focusing on specific targets, which we suspect may have been caused by the MAB condition. However, we were unable to confirm this within the current experiment. A definitive link between the MAB condition and avoidance behavior could be valuable for future research, particularly in contexts where guiding users away from certain scenes is desired.

Looking ahead, we aim to assess the effectiveness of ocularity singletons as a gaze-guiding feature in more dynamic and interactive VR environments. Additionally, we plan to investigate the applicability of this method in augmented reality (AR) and the unique challenges of integrating imperceptible gaze guidance into real-world settings.

\subsection{Privacy and Ethics}
Although this system is still in its early prototype phase, the results presented in this paper suggest that it can override normal user behavior at a subconscious level by subtly influencing gaze direction. This raises potential ethical concerns, particularly as the technology could guide user's attention without their conscious awareness. Future development of this system should consider privacy and user consent to ensure that the technology is developed in a responsible and transparent manner. 

\section{Conclusion}
In this paper, we explored the potential of using ocularity singletons as a novel method for imperceptible gaze guidance in virtual reality. By leveraging the principles of the V1 Saliency Hypothesis, we aimed to create a form of gaze guidance that operates without making perceptible changes to the visual environment, thereby preserving user immersion while effectively guiding attention. Our experimental results demonstrated that ocularity singletons, which rely on the disparity of individual eye inputs, can significantly influence visual search performance. Participants were able to detect targets more efficiently when the target itself is an ocularity singleton, and less efficiently when a random distractor is an ocularity singleton, supporting the hypothesis that V1-driven saliency can guide attention reflexively, even without conscious awareness. These findings suggest that ocularity singletons can serve as a powerful tool for subtle and non-intrusive gaze guidance in VR environments.

This imperceptible guidance technique opens up new possibilities for applications across various fields. In educational VR environments, imperceptible gaze guidance could enhance learning experiences by directing students' attention to critical content without distracting them from the immersive setting. Similarly, in medical and surgical training, where focus and immersion are essential, this technique could subtly guide trainees toward key areas in a procedure. Therapeutic and rehabilitation contexts might also benefit, as imperceptible guidance could support attention management in VR-based treatments without overtly controlling the user’s visual experience. In gaming and storytelling, this technique has the potential to lead players naturally through narratives and points of interest, preserving autonomy and engagement. Finally, in training simulations for hazardous situations, imperceptible guidance could subtly direct attention to critical objects or escape routes, enhancing realism and decision-making skills.

In conclusion, our findings lay the groundwork for a new approach in gaze guidance, one that seamlessly integrates with VR and potentially AR systems to enhance user experience without compromising immersion. Future work should focus on validating these results in more complex VR scenarios and exploring practical applications in real-world settings, ultimately advancing the possibilities for non-intrusive user attention management in extended reality technologies.

\section*{Acknowledgement}
Special thanks to Dr. Nora Castner and Dr. Marcus Nyström for their feedback and suggestions.

\bibliographystyle{ACM-Reference-Format}
\bibliography{references}

\end{document}